\documentclass[aps,twocolumn,showpacs,preprintnumbers,nofootinbib,tightenlines,useAMS]{revtex4}
\usepackage{bm}% bold math
\usepackage{times}
\usepackage{psfig}
\begin{document}

\title{Charged polytropic compact stars.}

\author{Subharthi Ray\footnote{present address : IUCAA, Pune 411007, India}, Manuel Malheiro}

\address{Instituto de F\'{\i}sica, Universidade Federal Fluminense,
Nit\'eroi 24210-340, RJ, Brazil}

\author{Jos\'e P. S. Lemos}

\address{Centro Multidisciplinar de Astrof\'{\i}sica - CENTRA, Departamento
de F\'{\i}sica,
Instituto Superior T\'ecnico, Av. Rovisco Pais 1, 1096 Lisboa, Portugal}

\author{Vilson T. Zanchin}

\address{Universidade Federal Santa Maria, Departamento de Fisica, BR-97119900,
Santa Maria, RS, Brazil}

\begin{abstract}
{In this work, we analyze the effect of charge in compact stars
considering the limit of the maximum amount of charge they can
hold. We find that the global balance of the forces allows a huge
charge ($\sim 10^{20}$ Coulomb) to be present in a neutron star
producing a very high  electric field ($\sim 10^{21}$ V/m). We
have studied the particular case of a polytropic equation of
state and assumed that the charge distribution is proportional to
the mass density. The charged stars have large mass and radius as
we should expect due to the effect of the repulsive Coulomb force
with the M/R ratio increasing with charge. In the limit of the
maximum charge the mass goes up to $\sim 10$ M$_\odot$ which is
much higher than the maximum mass allowed for a neutral compact
star. However, the local effect of the forces experienced by a
single charged particle, makes it to discharge quickly. This
creates a global force imbalance and the system collapses to a
charged black hole.}

\end{abstract}

\maketitle

\section{Introduction}

In 1924, Rosseland\cite{ross24} studied the possibility of a
self  gravitating  star  on   Eddington's  theory  to  contain  a  net
charge where the star  is modeled by a  ball of hot ionized gas (see also
Eddington \cite{ed26}).  In such a
system  the electrons  (lighter particles)  tend  to rise  to the  top
because  of  the  difference  in  the partial  pressure  of  electrons
compared to that of ions (heavier particles).  The motion of electrons
to the top and further escape from the star is stopped by the electric
field created  by the charge  separation. The equilibrium  is attained
after some  amount of electrons escaped leaving  behind an electrified
star  whose net positive  charge is  of about  100 Coulomb  per solar
mass, and building an interstellar  gas with a net negative charge. As
shown by Bally  and Harrison \cite{bally78}, this result  applies to any bound
system whose size is smaller  than the Debye length of the surrounding
media. The  conclusion is that a  star formed by  an initially neutral
gas cannot acquire a net  electric charge larger than about $100$C per
solar mass. It is expected that the sun holds some amount of
net charge due to the much more frequent escape of electrons than
that  of protons.  Moreover, it is also expected that  the escape
would stop when the electrostatic  energy of an electron $e\Phi$ is of
the order  of its thermal  energy $kT$. This  gives for a ball  of hot
matter with the  sun radius, a net charge  $Q \sim 6.7\times 10^{-6}T$
(in Coulomb). Hence, the escape  effect cannot lead to a net electric
charge much larger than a few hundred Coulomb for most of the gaseous
stars.

For Newtonian stars, the net charge of 100 C per solar mass is obtained
by the balance between the  electrostatic  energy  $eQ/r$ and
the gravitational  energy $mM/r$
(Glendenning\cite{glen00}). However, for very compact
stars, the high density and the relativistic effects  must be taken
into account \cite{bek71}.  In  a strong  gravitational  field, the
general relativistic effects  are felt and the star  needs more charge
to be in equilibrium.
Moreover, for  very compact stars,  the induced electric field  can be
substantially higher  than in the case  of the sun.  For instance, the
same amount of charge yields an electric field approximately $10^{9}$
times larger at  the surface of a neutron star than  at the surface of
the sun.  So, even a relatively  small amount of  net charge on
compact  stars can induce  intense electric  fields whose  effects may
become  important to  the structure  of the  star. This  fact deserves
further investigation.

The general relativistic analog for charged dust stars was
discovered by Majumdar \cite{maj47} and by Papapetrou
\cite{papa47}, and further discussed by Bonnor \cite{bon75} and
several other authors\cite{ivanov02c}. Study for  the stability
of charged fluid  spheres have been  done by
Bekenstein\cite{bek71}, Zhang et al.\cite{zhang82},  de Felice \&
Yu\cite{fel95}, Yu \& Liu\cite{yu00},  de Felice  et
al.\cite{fel99},  Anninos \& Rothman\cite{ani01} and others.
This  was indirectly  verified by Zhang et al.\cite{zhang82} who
found  that the structure of a neutron star, for a degenerate
relativistic fermi gas, is significantly  affected by the
electric charge  just when  the charge density  is close to  the
mass  density (in  geometric units).  In the investigations by
de  Felice et al., and by  Anninos \& Rothman, they assumed that
the charge  distribution followed particular functions of the
radial coordinate, and they  were mostly interested in the extreme
$Q=M$ case.

Our basic consideration to incorporate charge into  the system is
in the form of trapped  charged particles  where the  charge
goes  with  the positive value. The effect of charge does not
depend on its sign by our formulation. The energy density  which
appears from the electrostatic field will {\it add up} to the
total energy density of the system, which in turn  will  help  in
the  {\it  gaining}  of  the  total mass  of  the system. The
modified Tolman-Oppenheimer-Volkoff (TOV) equation now has extra
terms due to the presence of  the Maxwell-Einstein  stress
tensor. We solve the modified TOV equation for  polytropic
equation of state (EOS) assuming  that the  charge density goes
with the  matter density and discuss the  results. The formation
of this extra charge inside the star is however left open. A
mechanism to generate charge asymmetry for charged black holes
has been suggested recently by Mosquera Cuesta et al.
\cite{herman03} and the same may be applied for compact stars too
.

This article is arranged in the following way. In Section
\ref{sec:rel-form}, we show the basic formalism for the modified
TOV. In Section \ref{sec:poly}, we used this modified TOV on a
polytropic EOS, discuss the results and the stability of the
charged stars. Finally we make our conclusions in Section
\ref{sec:conc}.

\section{The modified Hydrostatic Equilibrium Equation}\label{sec:rel-form}

We take the metric for our static spherical star as

\begin{equation}
ds^2=e^\nu c^2 dt^2 - e^\lambda dr^2 - r^2(d\theta^2 + sin^2\theta d\phi^2).
\label{eq:1}
\end{equation}
The stress tensor $T^\mu_\nu$ will include the terms from the Maxwell's
equation and the
complete form of the Einstein-Maxwell stress tensor will be :
\begin{equation}
T^\mu_\nu = (P + \epsilon)u^\mu u_\nu + P \delta^\mu_\nu + \frac{1}{4\pi}
\left(F^{\mu\alpha}F_{\alpha\nu} - \frac{1}{4}\delta^\mu_\nu
F_{\alpha\beta} F^{\alpha\beta}\right)
\label{eq:tmunu}
\end{equation}
where  P is  the pressure,  $\epsilon$ is  the energy  density (=$\rho
c^2$) and  $u$-s are the  4-velocity vectors. For the  time component,
one  easily  sees  that  $u_t  = e^{-\nu/2}$  and  hence  $u^tu_t=-1.$
Consequently, the other components  (radial and spherical) of the four
vector are absent.

Now, the electromagnetic field is taken from the Maxwell's field equations
and hence they
will follow the relation
\begin{equation}
\left[ \sqrt{-g} F^{\mu\nu}\right]_{,\nu}~=~4 \pi j^\mu \sqrt{-g}
\label{eq:fmunu}
\end{equation}
where $j^\mu$ is the four-current density. Since the present choice of
the  electromagnetic  field  is  only  due to  charge,  we  have  only
$F^{01}=-F^{10}$, and the other terms  are absent.  In general, we can
derive  the   electromagnetic  field  tensor   $F^{\mu\nu}$  from  the
four-potential  $A_\mu$.  So,  for  non vanishing  field  tensor,  the
surviving  potential  is  $A_0=\phi$.  We  also  considered  that  the
potential has a spherical symmetry, i.e., $\phi=\phi(r)$.

The nonvanishing term in Eq.(\ref{eq:fmunu}) is when $\nu$=r. This gives
the electric field for both the $t$ and $r$ components as :
\begin{eqnarray}
\nonumber
\frac{1}{4\pi}
\left(F^{\mu\alpha}F_{\alpha\nu} - \frac{1}{4}\delta^\mu_\nu F_{\alpha\beta}
F^{\alpha\beta}\right)=\frac{-{\cal U}^2}{8\pi}
\end{eqnarray}
where,
\begin{equation}
{\cal U}(r)=\frac{1}{r^2}\int^r_0 4 \pi r^2 \rho_{ch} e^{\lambda/2} dr.
\label{eq:ucal2}
\end{equation}
is the electric field. So, the total charge of the system is
\begin{equation}
Q=\int^R_0 4 \pi r^2 \rho_{ch} e^{\lambda/2} dr
\label{eq:charge}
\end{equation}
where R is the radius of the star.

The mass of the star is now due to the total contribution
of  the energy  density of the  matter and  the electric  energy ($\frac{{\cal
U}^2}{8\pi}$) density. The mass takes the new form as
\begin{equation}
M_{tot}(r)=\int^r_04\pi r^2\left(\frac{\epsilon}{c^2}+\frac{{\cal
U}^2}{8\pi c^2}\right)dr
\label{eq:mnew}
\end{equation}
and the metric coefficient is given by
\begin{equation}
e^{-\lambda}=1-\frac{2GM_{tot}(r)}{c^2r}.
\label{eq:lamnew}
\end{equation}

The  stress   tensor is conserved (${T^\mu_\nu}_{,\mu}=0$).
Hence, one  gets   the  form   of   the  hydrostatic equation from
it as :
\begin{eqnarray}
\nonumber
\frac{dP}{dr}&=&-\frac{G\left[M_{tot}(r)+4\pi
r^3\left(\frac{P}{c^2}-\frac{{\cal U}^2}{8\pi c^2}
\right)\right](\epsilon+P)}{c^2r^2\left(1-\frac{2GM_{tot}}{c^2r}\right)}\\
&&+\rho_{ch}{\cal U}e^{\frac{\lambda}{2}}.
\label{eq:dpdr}
\end{eqnarray}
We solve the Eqns. (\ref{eq:mnew}, \ref{eq:lamnew} \& \ref{eq:dpdr})
simultaneously to get
our results for the charged compact stars.

\section{Effect of charge on polytropic stars}\label{sec:poly}
%\subsection{Mass radius relations and structure}

We study here the effect of charge on a model independent polytropic EOS.
We assume the  charge is proportional to
the mass density ($\epsilon$) like
\begin{equation}
\rho_{ch}=f \times\epsilon
\end{equation}
where $\epsilon=\rho c^2$ is in [MeV/fm$^3$]. In geometrical
units, this can be written as
\begin{equation}
\rho_{ch}=\alpha \times\rho
\label{eq:no-dim}
\end{equation}
where charge is expressed in units of mass and charge density in units of
mass density.
This $\alpha$ is related to our charge fraction $f$ as
\begin{equation}
\alpha=f\times\frac{0.224536}{\sqrt{G}}=f\times 0.86924 \times 10^3.
\label{eq:no-dim-conv}
\end{equation}
Our choice of charge distribution  is a reasonable assumption
in the sense that large mass can hold large amount of charge.

The polytropic EOS is given by
\begin{equation}
P=\kappa \rho^{1+1/n}
\end{equation}
where  $n$ is the  polytropic index  and is  related to  the
exponent $\Gamma$  as  $\Gamma=1+ \frac{1}{n}$. In  the
relativistic regime,   the  allowed   value   of  $\Gamma$   is
$\frac{4}{3}$   to $\frac{5}{3}$.   We   have    considered the
adiabatic   case   of $\Gamma=\frac{5}{3}$   and   the
corresponding   value   of  $n$   is 1.5. Primarily,  our  units
of  matter  density  and  pressure are  in MeV/fm$^3$.   We
chose   a   value   of   $\kappa$   as 0.05 [fm]$^{8/3}$ for our
polytrope  that reproduces quite well realistic EOS for neutron
stars \cite{taurines2001}. It should be noted that the amount of
charge we find implies a very small ratio $Z/A\simeq 10^{-18}$
which justifies to use an EOS which is calculated for neutral
matter.

\begin{figure}[htbp]
\centerline{\psfig{figure=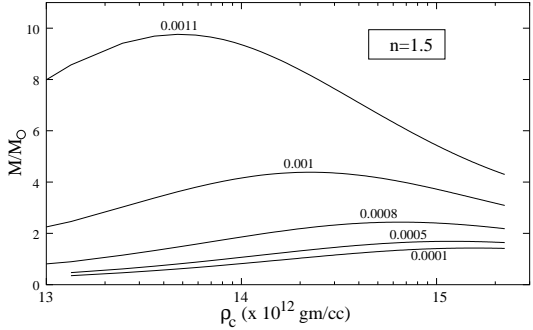,width=7cm}}
\caption{Central density against mass for different values of the factor $f$.}
\label{fig:poly-e-m}
\end{figure}

In  Fig.(\ref{fig:poly-e-m}), we plot the mass as function of the
central density, for  different values  of the  charge fraction
$f$. For the charge fraction $f=0.0001$, we do not see any
departure on the stellar structure from that of the chargeless
case. This value of $f$ is $critical$ because any increase in the
value beyond this, shows enormous effect on the structure. The
increase of the maximum mass of the star is very much non-linear,
as can be seen from the Fig. (\ref{fig:poly-e-m}).

\begin{figure}[htbp]
\centerline{\psfig{figure=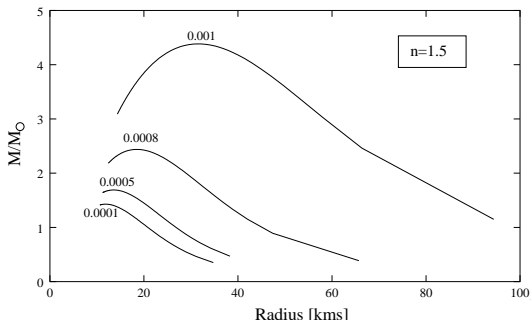,width=7cm}}
\caption{Mass as a function of radius, for different values of the factor $f$.}
\label{fig:poly-m-r}
\end{figure}

In Fig.(\ref{fig:poly-m-r}) we plotted the mass-radius relation.
Due to the effect of the repulsive force, the charged stars have
large radius and larger mass as we should expect. Even if the
radius is  increasing with  the mass,  the M/R ratio is  also
increasing, but much slower.  For the lower charge fractions,
this increase in the radius is very small, but a  look at the
structure  for the fraction  $f=0.001$ reveals that for a mass of
4.3 M$_\odot$, the radius goes as high as 35 km. Though the
compactness of the stars are  retained, they are now better to
be  called as  $charged~compact~stars$ rather  than
$charged~neutron~ stars$. The charge fraction in the limiting
case of maximally allowed value goes up to $f=0.0011$, for which
the maximum mass stable star forms at a lower central density
even smaller than the nuclear matter density. This extreme case
is not shown in Fig.(\ref{fig:poly-m-r}) because the radius of
the star and its mass is very high (68 km \& 9.7 M$_\odot$
respectively) which suppresses the curves of the lower charge
fractions due to scaling. For this star, the mass contribution
from the electric energy density is 10\% than that from the mass
density. It can be checked by using relation
(\ref{eq:no-dim-conv}) that this charge fraction $f=0.0011$
corresponds to $\rho_{ch}=0.95616\times\rho$ in geometrical units.

\begin{figure}[htbp]
\centerline{\psfig{figure=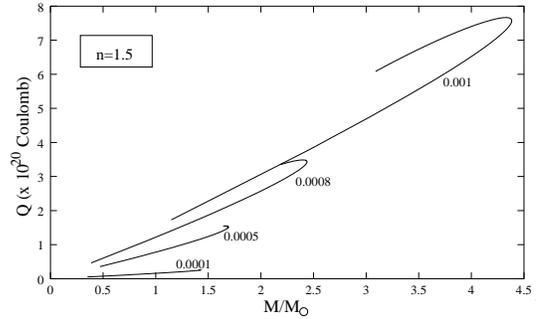,width=7cm}}
\caption{\label{fig:poly-q-m}The variation of the charge with mass for
different $f$.}
\end{figure}

The Q$\times$ M diagram of Fig.(\ref{fig:poly-q-m}) shows the
mass of the stars against their surface charge. We have made the
charge density proportional to  the energy  density and so  it
was expected  that the charge, which is  a volume integral of the
charge  density, will go in the same  way as the  mass, which is
also a volume integral  over the mass density.  The slope of the
curves comes from the different charge fractions.The nature of
the curves  in fact reflects that charge varies with mass  (with
the  turning back  of the curves  all falling  in the `unstable
zone' and is  not taken into consideration). If we consider that
the maximum allowed charge estimated  by the condition (${\cal U}
\simeq \sqrt{8\pi P} < \sqrt{8\pi\epsilon}$) for $\frac{dP}{dr}$
to be negative  (Eq.~(\ref{eq:dpdr})),  we see that the  curve
for the maximum charge  in Fig.(\ref{fig:poly-q-m}) has a  slope
of 1:1 (in a  charge scale of 10$^{20}$  Coulombs\footnote{As
electric energy density and pressure needs to be of the same
order, so from the fine structure constant $\alpha =
\frac{e^2}{\hbar c} =\frac{1}{137},$ we get a relation for the
charge and MeV/fm$^3$ which comes out as $1 C \simeq 0.75 \times
10^{19} [MeV fm]^{1/2}.$}). This scale can easily be understood
if we write the charge as
$Q=\sqrt{G}M_\odot\frac{M}{M_\odot}\simeq10^{20}\frac{M}{M_\odot}{\rm
Coulombs}.$ This charge Q  is the charge at  the surface of the
star where the  pressure and  also $\frac{dP}{dr}$  are zero. So,
at  the surface, the  Coulomb force is essentially balanced  by
the gravitational  force and the  relation of the charge and mass
distribution we  found is exactly the same for the case of charged
dust sphere  discussed  earlier  by  Papapetrou\cite{papa47} and
Bonnor\cite{bon75}.

The total  mass of the system  $\rm M_{tot}$ increases with
increasing charge because the  electric energy density $adds~on$
to  the mass energy density. This change in  the mass is low for
smaller charge  fraction  and going  up to  7 times  the value  of
chargeless case  for maximum allowed charge  fraction $f=0.0011$.
The most effective term   in  Eq.(\ref{eq:dpdr})  is  the factor
($\rm M_{tot}+4\pi  r^3P^*$). $P^*= P-\frac{{\cal U}^2}{8\pi}$ is
the effective pressure of the system because the effect of charge
decreases the outward fluid pressure, negative in sign to the
inward gravitational pressure. With the increase of charge, the
value of P$^*$ decreases, and hence the gravitational negative
part  of Eq.(\ref{eq:dpdr}) decreases. So, with the softening of
the pressure  gradient, the system allows more radius for the
star until  it reaches  the surface where the pressure (and
$\frac{dP}{dr}$) goes to zero. We should stress that because
$\frac{{\cal U}^2}{8\pi}$ cannot be  too much larger than the
pressure in order to maintain  $\frac{dP}{dr}$  negative as
discussed before,  so   we  have  a  limit  on  the charge, which
comes from the relativistic effects of the gravitational force
and not just only from the repulsive Coulombian part.

\begin{figure}[htb]
\centerline{\psfig{figure=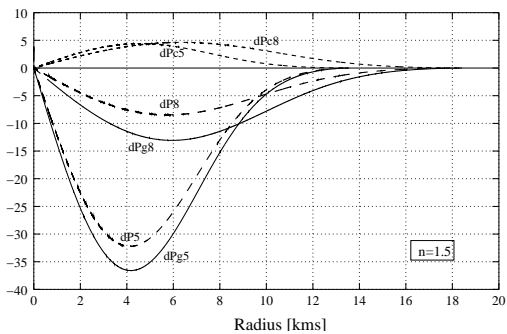,width=7cm}} \caption{The
positive Coulomb part and the gravitational negative part of the
pressure gradient together with the total ($\frac{dP}{dr}$) are
shown here for two  different values of  the charge factor $f$.
%(=0.0005 \& 0.0008)
For $f=0.0005$ and  0.0008 coming from the
$matter~part$ are denoted as dPg5 and dPg8 respectively, those
from $Coulomb~part$ are dPc5 and dPc8 respectively. The
corresponding totals are dP5 and dP8.} \label{fig:poly-dpdr}
\end{figure}
This effect  is shown  in Fig.(\ref{fig:poly-dpdr})  where we
have plotted  both the positive  Coulomb part and the  negative matter
part  of the  pressure gradient.  The plots are for  two values of  the
charge fraction
$f=0.0005$  and  $f=0.0008$.  The  positive  part  of  $\frac{dP}{dr}$
maintains its almost constant value because the charge fraction $f$ is
the controller  of the same,  and in our  case, they differ by  a very
small percentage.  In the negative  part, the changes are  drastic and
are mainly brought by the effective pressure as we already discussed.

In our high density system, the gravitational and the Coulomb
forces are highly coupled. Although it is difficult to disentangle
the forces, but to a common belief, it can be considered that the
charged particles, due to their self created huge field, will
leave the star very soon. This process will however lead to an
imbalance of the global forces acting on the star, which were
previously balanced by the Gravitational and the Coulomb forces.
This process will help in the star to further collapse to a
charged black hole. We say it charged because, by the time the
system collapses, all the charge has not left the star, and they
get trapped inside the black hole \cite{sray03}.

\section{Conclusions}\label{sec:conc}
In our study, we have shown that a high density system like a neutron star
can hold huge charge of the order of 10$^{20}$ Coulomb considering the
global balance of forces. With the increase of charge, the maximum mass
of the star recedes back to a lower density regime. The stellar mass
also increases rapidly in the critical limit of the maximum charge
content, the systems can hold. The radius also increases accordingly,
however keeping the M/R ratio increasing with charge. The increase in mass
is primarily brought in by the softening of the pressure gradient due
to the presence of a Coulombian term coupled with the Gravitational matter
part.
Another intrinsic increase in the mass term comes through the addition of
the electric energy density to the mass density of the system.

The inside electric field of the charged stars are very high and
crosses the critical field limit for pair creation (Bekenstein
\cite{bek71}). However, this issue is debatable because the
critical field has been calculated for vacuum and one does not
really know what the value will be in a high density system. The
stability of the charged stars are however ruled out from the
consideration of forces acting on individual charged particles.
They face enormous radial repulsive force and leave the star in a
very short time. This creates an imbalance of forces and the
gravitational force overwhelms the repulsive Coulomb and fluid
pressure forces and the star collapses to a charged black hole.

Finally, these charged stars are supposed to be very short lived,
and are the intermediate state between a supernova collapse and
charged black holes.

SR acknowledges the FAPERJ for research support, MM for the partial CNPQ
support, and JSPL and VTZ for the hospitality of Observat\'{o}rio Nacional,
Rio de Janeiro.

\end{document}